\documentclass[reqno,dvips,12pt]{article}
\usepackage{amsmath,amssymb,amsfonts}
\newcommand{\ol}{\overline}
\newcommand{\ov}{\overline}
\newcommand{\wt}{\widetilde}
\newcommand{\cH}{{\cal H}}
\newcommand{\Symm}{\mathop{\rm Symm}}
\newcommand{\Asymm}{\mathop{\rm Asymm}}
\newcommand{\Sp}{\mathop{\rm Sp}\nolimits}
\newcommand{\nn}{\nonumber}

\newcommand{\cF}{{\cal F}}
\newcommand{\bR}{{\bf R}}
\newcommand{\bT}{{\bf T}}
\newcommand{\wh}{\widehat}
\newcommand{\str}{\stackrel}
\newcommand{\pa}{\partial}

\renewcommand{\r}{\right)}
\renewcommand{\l}{\left(}

\begin{document}
\title {On the superfluidity of classical liquid in nanotubes}
\author{V.P.Maslov}
\date{}
\maketitle

\begin{abstract}
In 2001, the author proposed the ultra second quantization method.
The ultra second quantization of the Schr\"odinger equation,
as well as its ordinary second quantization, is a representation of the $N$-particle
Schr\"odinger equation, and this means that basically the ultra second quantization
of the  equation is the same as the original $N$-particle equation:
they coincide in $3N$-dimensional space.

We consider a short action pairwise potential $V(x_i -x_j)$. This means that
as the number of particles tends to infinity, $N\to\infty$,
interaction is possible for only a finite number of particles.
Therefore, the potential depends on $N$ in the following way:
$V_N=V((x_i-x_j)N^{1/3})$. If $V(y)$ is finite with support
$\Omega_V$, then as  $N\to\infty$ the support engulfs a finite number of particles,
and this number does not depend on $N$.

As a result, it turns out that the superfluidity occurs
for velocities less than $\min(\lambda_{\text{crit}}, \frac{h}{2mR})$,
where $\lambda_{\text{crit}}$ is the critical Landau velocity
and $R$ is the radius of the nanotube.
\end{abstract}

\textbf{1.}  In order to distinguish the classical theory in its modern understanding
from the quantum theory, it is necessary to modify (somewhat) the ideology habitual
to physicists, for whom the classical theory is simply the whole body of physics
as it existed in the 19th century before the appearance of quantum theory.
Actually, the correct meaning is that the classical theory is
the limit of the quantum one as $h\to 0$.

Thus, Feynman correctly understood that spin is a notion of classical mechanics. Indeed,
it is obtained via a rigorous passage from quantum mechanics to classical mechanics
\cite{TeorVoz}. In a similar same way, the polarization of light does not disappear
when the frequency is increased, and is therefore a property of geometric rather
than wave optics, contrary to the generally accepted belief,
which arose because the polarization of light was discovered as the result
of the appearance of wave optics.

Consider a ``Lifshits hole'', i.e. a one-dimensional Schr\"odinger equation with potential symmetric with respect to the origin of coordinates with two troughs.
Its eigenfunctions are symmetric or antisymmetric with respect to the origin. As
$h\to 0$ this symmetry remains, and since the square of the modulus of the eigenfunction corresponds to the probability of the particle to remain in the troughs, it follows that in the limit as $h\to 0$, i.e., in the ``classical theory", for energies less than those required to pass over the barrier, the particle is simultaneously located in two troughs, although a classical particle cannot pass through the barrier. Nevertheless, this simple example shows how the ideology of the ``classical theory'' must be modified.

To understand this paradox, one must take into consideration the fact that the symmetry must be very precise, up to ``atomic precision'', and that stationary state means  a state that arises in the limit for ``infinitely long''  time.

When we deal with nanotubes whose width is characterized by ``atomic'' or ``quantum'' dimensions, then new unexpected effects occur in the ``classical'' theory. Thus, already in 1958  \cite{DAN_58}, I discovered a strange effect of the standing longitudinal wave type in a slightly bent infinite narrow tube, for the case in which its radius is the same everywhere with atomic precision. It was  was impossible at the time to implement this effect in practice, which would have allowed to obtain a unimode laser, despite   A.M.Prokhorov's great interest in the effect.

\textbf{2.} Now let us discuss the notion known as ``collective oscillations''   in classical physics and as ``quasiparticles''  in quantum physics. In classical physics, it is described by the Vlasov equation for selfcompatible (or mean) fields, in quantum physics, by the Hartrey (or the Hartrey-Fock) equation.

(1) Variational equations depend on where (i.e., near what solutions of the original equation) we consider the variations. For example, in \cite{QuasiPart5, QuasiPart6, QuasiPart7} we considered variations near a microcanonical distribution in an ergodic construction, while in \cite{QuasiPart1, QuasiPart2, QuasiPart3, QuasiPart4} this was done near a nanocanonical distribution concentrated on an invariant manifold of lesser dimension, i.e., not on a manifold of constant energy but, for example, on a Lagrangian manifold of dimension coinciding with that of the configuration space.

(2) Let us note the following crucial circumstance. The solution of the variational equation for the Vlasov equation   \textit{does not coincide} with the classical limit for variational equations for the mean field equations in  quantum theory.

Consider the mean field equation in the form

\begin{equation}\label{Hart}
ih\frac{\pa}{\pa t}\varphi^t (x)= \bigg(-\frac{h^2}{2m}\Delta
+W_t(x)\bigg) \varphi^t(x), \qquad W_t(x)=U(x)+\int
V(x,y)|\varphi^t(y)|^2dy,
\end{equation}
with the initial condition
 $\varphi|_{t=0}=\varphi_0$, where $\varphi_0$
satisfies $\varphi_0 \in W_2^\infty (\bR^\nu), \int
dx|\varphi_0(x)|^2=1$.

In order to obtain asymptotics of the complex germ type \cite{VKB}
one must write out the system consisting of the Hartrey equation and its dual,
then consider the corresponding variational equation, and, finally, replace
the variations  $\delta\varphi$ and $\delta\varphi^*$ by the independent functions $F$ and $G$. For the functions $F$ and $G$, we obtain the following system of equations:

\begin{eqnarray}\label{Shv1}
&& \ i\frac{\pa F^t(x)}{\pa t} = \int dy
\bigg(\frac{\delta^2H}{\delta\varphi^*(x)\delta\varphi(y)}F^t(y)+
\frac{\delta^2H}{\delta\varphi^*(x)\delta^*\varphi(y)}G^t(y)\bigg);
\\
&&-i\frac{\pa G^t(x)}{\pa t} = \int dy
\bigg(\frac{\delta^2H}{\delta\varphi(x)\delta\varphi(y)}F^t(y)+
\frac{\delta^2H}{\delta\varphi(x)\delta^*\varphi(y)}G^t(y)\bigg).
\nn
\end{eqnarray}

The classical equations are obtained from the quantum ones, roughly speaking,
by means of a substitution of the form $\varphi= \chi e^{\frac ih S}$
(the VKB method), $\varphi^*= \chi^* e^{\frac ih S^*}, where \ S=S^*, \
\chi=\chi(x,t)\in C^\infty, \ S=S(x,t)\in C^\infty$.

To obtain the variational equations, it is natural to take the variation not only of the
limit equation for  $\chi$ and  $\chi^*$, but also for the functions $S$ and
$S^*$. This yields a new important term of the equation for collective oscillations.

Let us describe this fact for the simplest example, which was studied in
N.N.Bogolyubov's famous paper concerning ``weakly ideal Bose gas" \cite{Bogol}.

Suppose $U=0$ in equation~(\ref{Hart}) in a three-dimensional cubical
box of edge $L$, the wave functions satisfying the periodicity condition (i.e., the
problem being defined on the 3-torus with generators of lengths  $L, L, L$).
Then the function

\begin{equation}\label{bog1}
  \varphi(x)=L^{-3/2} e^{i/h(px-\Omega t)},
\end{equation}
where $p=2\pi n/L$, $n$ -- is an integer,  satisfies the equation
(\ref{Hart}) for
\begin{equation}\label{bog2}
  \Omega=\frac{p^2}{2m}+L^{-3}\int dx V(x).
\end{equation}

Consider the functions $F^{(\lambda)}(x)$ and $G^{(\lambda)}(x)$, where
$\lambda=2\pi {n}/L, {n}\neq 0$, ${n}$ is an integer, given by

\begin{eqnarray}\label{bog3}
&& F^{(\lambda)t}(x)= L^{-3/2}\rho_\lambda e^{\frac
ih|(p+\lambda)x +(\beta-\Omega)t|}, \nn \\
&& G^{(\lambda)t}(x)= L^{-3/2}\sigma_\lambda e^{\frac
ih|(-p+\lambda)x +(\beta+\Omega)t|};
\end{eqnarray}
here
\begin{eqnarray}\label{bog4}
&&-\beta_\lambda\rho_\lambda=\bigg(\frac{(p+\lambda)^2}{2m}-\frac{p^2}{2m}
+\wt{V}_\lambda\bigg)\rho_\lambda+V_\lambda\sigma_\lambda, \nn \\
&&\beta_\lambda\rho_\lambda=\bigg(\frac{(p-\lambda)^2}{2m}-\frac{p^2}{2m}
+\wt{V}_\lambda\bigg)\sigma_\lambda+V_\lambda\rho_\lambda,\\
&&|\sigma_\lambda|^2 -|\rho_\lambda|^2=1, \qquad
\wt{V}_\lambda=L^{-3}\int dx V(x)e^{\frac ih \lambda x}. \nn
\end{eqnarray}

From the system (\ref{bog4}), we find
\begin{equation}\label{bog5}
\beta_\lambda = -p \lambda+
\sqrt{\bigg(\frac{\lambda^2}{2m}+\wt{V}_\lambda\bigg)^2-\wt{V}_\lambda^2}.
\end{equation}

In this example $u=e^{\frac ih s(x,t)}, \ u^*=e^{-\frac
{s(x,t)}{h}}$, where $s(x,t,)= px+\beta t$, while the variation of the action for the vector
 $\big({\delta u}, {\delta u^*}\big)$ equals $\lambda x\pm\Omega t$.

Under a more accurate passage to the limit, we obtain $\wt{V}_\lambda \to V_0
=L^{-3} \int dx V(x)$.

Thus, in the classical limit, we have obtained the famous Bogolyubov relation
 (\ref{bog5}).  In the case under consideration
$u(x)=0$ and, as in the linear Schr\"odinger equation, the exact solution coincides with the quasiclassical one. In the paper  \cite{QuasiPart4}, the case
$u(x)\neq 0$ is investigated, and it turns out that the relation similar to
 (\ref{bog5}) is the classical limit as $h\to 0$ of the variational equation in this general case. The curve showing the dependence of $\beta_\lambda$ on $\lambda$
is known as the \textit{Landau curve} and determines the superfluid state. The value
 $\lambda_{\text{cr}}$ for which superfluidity disappears is called the \textit{Landau critical level}. Bogolyubov explains the superfluidity phenomenon in the following terms:
``the `degenerate condensate' can move without friction relatively to elementary perturbations with any sufficiently small velocity''  \cite{QuasiPart4}, p.~210.

However, there is no Bose-Einstein condensate whatever in these mathematical considerations, it is just that the spectrum defined for
$\lambda <\lambda_{\text{cr}}$  is a positive spectrum of quasiparticles. This means it is  metastable (see \cite{Masl_Shved}). The Bose-Einstein condensate is not involved here, it is only needed only to show that it would be wrong to believe  that this argument works for a classical liquid, as one might think from the considerations above.

Indeed, for example, the molecules of a classical nondischarged liquid are, as a rule, Bose particles. For such a liquid, one can write out the $N$-particle equation, having in mind that each particle (molecule) is neutral and consists of an even number $l$ of neutrons. Thus each $i$th particle is  a point in $3(2k+l)$-dimensional space, where
$k$ is the number of electrons, $x_i \in R^{6k+3l}$, depends on the potential $u(x_i), \ x_i \in R^{6k+3l}$ and we can consider the  $N$-particle equation for $x_i, i=1, \dots, N$,
with pairwise interaction $V(x_i-x_j)$.

\textbf{3.} However, there is a purely mathematical explanation of this paradox. The thing is that Bogolyubov found only one series of of ppoints in the spectrum of the many particle problem. Landau wrote ``N.N.Bogolyubov recently succeeded, by means of a clever application of second quantization, in finding the general form of the energy spectrum of a Bose-Einstein gas with weak interaction between the particles''
(\cite{landau}, p.~43).
But this series is not unique, i.e., the entire energy spectrum was not obtained.

In 2001, the author proposed the method of ultra second quantization \cite{Book_Ultravt}; see also
\cite{FAN_2000}, \cite{Uspehi_2000}, \cite{RJ_2001}, \cite{RJ_2001_2}, \cite{Mas1},

The ultra second quantization of the Schr\"odinger equation,
as well as its ordinary second quantization, is a representation of the $N$-particle
Schr\"odinger equation, and this means that basically the ultra second quantization
of the  equation is the same as the original $N$-particle equation:
they coincide in $3N$-dimen\-sional space. However, the replacement of the creation and annihilation  operators by $c$-numbers, in contrast with the case of second quantization, does not yield the correct asymptotics, but it turns out that it coincides with the result of applying the Schroeder variational principle or the Bogolyubov variational method.

For the exotic Bardin potential, the correct asymptotic solution coincides with the one obtained by applying the ultra second quantization method described above.
In the case of general potentials, in particular for pairwise interaction potentials, the answer is not the same. Specifically, the ultra second quantization method gives other asymptotic series of eigenvalues corresponding to the $N$-particle Schr\"odinger equation, and these eigenvalues, unlike the Bogolyubov  ones (7), are not metastable.

It turns out that the main point is not related to the Bose-Einstein condensate,
but has to do with the width of the capillary (the nanotube) through which the liquid flows. If we consider a liquid in a capillary or a nanotube of sufficiently small radius the velocity corresponding to metastable states is not small. Hence at smaller velocities the flow will be without friction.

The condition that the liquid does not flow through the boundary of the nanotube is a Dirichlet condition. It yields a standing wave, which can be regarded as a pair particle--antiparticle: a particle with momentum $p$ orthogonal to the boundary of the tube, and an antiparticle with momentum $-p$.

We consider a short action pairwise potential $V(x_i -x_j)$. This means that as the number of particles tends to infinity, $N\to\infty$, interaction is possible for only a finite number of particles. Therefore, the potential depends on $N$ in the following way:\linebreak $V_N=V((x_i-x_j)N^{1/3})$. If $V(y)$ is finite with support
$\Omega_V$, then as  $N\to\infty$ the support engulfs a finite number of particles, and this number does not depend on $N$.

As the result, it turns out that for velocities less than
$\min(\lambda_{\text{cr}}, \frac{h}{2mR})$, where $\lambda_{\text{cr}}$
is the critical Landau velocity and  $R$ is the radius of the nanotube, superfluidity occurs.

Now let me present my own considerations, which are not related to the mathematical exposition. Viscosity is due to the collision of particles: the higher the temperature, the greater the number of collisions. In a nanotube, there are few collisions, and only with the walls,
and those are taken into account by the author's series. It is precisely this circumstance, and not the Bose-Einstein condensate, which leads to the weakening of viscosity and so to superfluidity. What I am saying is that the main factor in the superfluidity phenomenon, even for liquid helium 4, is not the condensate, but the presence of an extremely thin capillary \cite{TMF_2005}, \cite{RJ_2005}.

\newpage

\section{Ultrasecondary quantization}
In the papers \cite{Book_Ultravt,1,2,3} the notion of ultrasecondary quantization was introduced for problems of quantum mechanics and statistical physics. Let us recall the notation and the main facts in the case of the quantization of pairs (particle--particle numbers) and for pairs consisting of two particles. Quantization of pairs will allow us to take into account pairwise correlations of particles in the construction of the asymptotics. The space of ultrasecondary quantization is  $\cF$, the Fock boson space \cite{7},  $\widehat{b}^+(x,s)$ is the creation operator of particles with number $s$,
$\widehat{b}^-(x,s)$ is the  annihilation  operator of particles with number  $s$
in the space $\cF$ \cite{7}, $\widehat{B}^+(x,x')$ is the creation operator of pairs particles, $\widehat{B}^-(x,x')$  is the annihilation operator of pairs of particles in this space. These operators satisfy the following commutation relations:

\begin{eqnarray}\label{bol1}
&&[\widehat{b}^-(x,s),\widehat{b}^+(x',s')]
=\delta_{ss'}\delta(x-x'),\qquad
[\widehat{b}^{\pm}(x,s),\widehat{b}^{\pm}(x',s')]=0,
\nn\\
&&[\widehat{B}^{-}(x_1,x_2)\widehat{B}^{+}(x'_1,x'_2)]
=\delta(x_1-x'_1)\delta(x_2-x'_2),
\nn\\
&&[\widehat{B}^\pm(x_1,x_2),\widehat{B}^\pm(x'_1,x'_2)]=0,
\\
&&[\widehat{b}^\pm(x,s),\widehat{B}^\pm(x'_1,x'_2)]
=[\widehat{b}^\pm(x,s),\widehat{b}^\mp(x'_1,x'_2)]=0. \nn
\end{eqnarray}

Further, $\Phi_0$ is the vacuum vector in the space $\cF$,
which possesses the following properties:

\begin{equation}\label{bol2}
\widehat{b}^-(x,s)\Phi_0=0,\qquad \widehat{B}^-(x_1,x_2)\Phi_0=0.
\end{equation}
The variable $x$ lies on the 3-dimensional torus of size $L\times L\times L$,
which we shall denote by $\bT$. The variable $s$ is discrete,
$s=0,1,\dots$; $s$ is called the number or the \textit{statistical spin}.  Any vector
$\Phi$ of the space $\cF$ can be uniquely represented in the form:

\begin{eqnarray}\label{bol3}
\Phi&=&\sum^{\infty}_{k=0}\sum^{\infty}_{M=0}\frac1{k!M!}
\sum^{\infty}_{s_1=0}\dots\sum^{\infty}_{s_k=0} \int\dots\int
dx_1\dots dx_k dy_1\dots dy_{2M}\times
\nn\\
&&\quad \times
\Phi_{k,M}(x_1,s_1;\dots;x_k,s_k;y_1,y_2;\dots;y_{2M-1},y_{2M})\times
\\
&&\quad \times
\widehat{b}^+(x_1,s_1)\cdot\dots\cdot\widehat{b}^+(x_k,s_k)
\widehat{B}^+(y_1,y_2)\cdot\dots\cdot
\widehat{B}^+(y_{2M-1},y_{2M})\Phi_0, \nn ,
\end{eqnarray}
where the function
$\Phi_{k,M}(x_1,s_1;\dots;x_k,s_k;y_1,y_2;\dots;y_{2M-1},y_{2M})$
is symmetric with respect to transpositions of pairs of variables  $(x_j,s_j)$ and
$(x_i,s_i)$ and symmetric with respect to transpositions of pairs of variables
$(y_{2j-1},y_{2j})$ and $(y_{2i-1},y_{2i})$.

In the boson case, we introduce the space  $\cF^{\rm Symm}_{k,M}$
consisting of the vectors $\Phi$ for which $\Phi_{k',M'}=0$ whenever $(k',M')\ne(k,M)$,
while $\Phi_{k,M}$ is a symmetric function of the variables
$x_1,x_2,\dots,x_k,y_1,y_2,\dots,y_{2M}$. In the fermion case we similarly introduce
the space $\cF^{\rm Asymm}_{k,M}$ consisting of the vectors $\Phi$ such that $\Phi_{k',M'}=0$ for $(k',M')\ne(k,M)$, and $\Phi_{k,M}$ is an antisymmetric function of the variables  $x_1,x_2,\dots,x_k,y_1,y_2,\dots,y_{2M}$. The orthogonal projector of the space  $\cF$ on the subspace $\cF^{\rm Symm}_{k,M}$ is of the form \cite{Book_Ultravt}-\cite{3}:
\begin{eqnarray} \label{bol4}
&&\widehat{\Pi}^{\rm Symm}_{k,M} =\frac1{k!M!}
\sum^{\infty}_{s_1=0}\dots\sum^{\infty}_{s_k=0} \int\dots\int
dx_1\dots dx_k dy_1\dots dy_{2M}\times
\nn\\
&&\quad \times \widehat{b}^+(x_1,s_1)\cdot\dots\cdot
\widehat{b}^+(x_k,s_k) \widehat{B}^+(y_1,y_2)\cdot\dots\cdot
\widehat{B}^+(y_{2M-1},y_{2M}) \times
\nn\\
&&\quad \times \Symm_{x_1\dots x_ky_1\dots y_{2M}}
\Big(\widehat{b}^-(x_1,s_1)\cdot\dots\cdot \widehat{b}^-(x_k,s_k)
\widehat{B}^-(y_1,y_2)\cdot\dots
\nn\\
&&\qquad\qquad\qquad\qquad\qquad\qquad \cdot
\widehat{B}^-(y_{2M-1},y_{2M})\Big)\times
\\
&&\quad \times \exp\bigg(-\sum^{\infty}_{s=0}\int dx
\widehat{b}^+(x,s)\widehat{b}^-(x,s) -\iint dy dy'
\widehat{B}^+(y,y')\widehat{B}^-(y,y')\bigg), \nn ,
\end{eqnarray}
where $\Symm_{x_1\dots x_ky_1\dots y_{2M}}$ is the symmetrization operator in the variables $x_1,\dots,x_k$, \linebreak $y_1,\dots,y_{2M}$,
while the operators $\widehat{b}^+(x,s)$, $\widehat{b}^-(x,s)$,
$\widehat{B}^+(y,y')$, $\widehat{B}^-(y,y')$ are ordered in the Vick way
~\cite{7}. The orthogonal projector of the space
$\cF$ on the subspace $\cF^{\rm Asymm}_{k,M}$ is of the form
\cite{Book_Ultravt}:
\begin{eqnarray}\label{bol5}
&&\widehat{\Pi}^{\rm Asymm}_{k,M} =\frac1{k!M!}
\sum^{\infty}_{s_1=0}\dots\sum^{\infty}_{s_k=0} \int\dots\int
dx_1\dots dx_k dy_1\dots dy_{2M}\times
\nn\\
&&\quad \times \widehat{b}^+(x_1,s_1)\cdot\dots\cdot
\widehat{b}^+(x_k,s_k) \widehat{B}^+(y_1,y_2)\cdot\dots\cdot
\widehat{B}^+(y_{2M-1},y_{2M})\times
\nn\\
&&\quad \times \Asymm_{x_1\dots x_ky_1\dots y_{2M}}
\Big(\widehat{b}^-(x_1,s_1)\cdot\dots\cdot \widehat{b}^-(x_k,s_k)
\widehat{B}^-(y_1,y_2)\cdot\dots
\nn\\
&&\qquad\qquad\qquad\qquad\qquad\qquad \cdot
\widehat{B}^-(y_{2M-1},y_{2M})\Big)\times \nn
\\
&&\quad \times\exp\bigg(-\sum^{\infty}_{s=0}\int dx
\widehat{b}^+(x,s)\widehat{b}^-(x,s) -\iint dy dy'
\widehat{B}^+(y,y')\widehat{B}^-(y,y')\bigg), \nn
\end{eqnarray}
where $\Asymm_{x_1\dots x_ky_1\dots y_{2M}}$ is the antisymmetrization
operator in the variables $x_1,\dots,x_k,$ \linebreak $y_1,\dots,y_{2M}$.
Here and everywhere else below, unless explicitely stated, the operators
$\widehat{b}^+(x,s)$, $\widehat{b}^-(x,s)$,
$\widehat{B}^+(y,y')$, $\widehat{B}^-(y,y')$ are ordered in the Vick way.

Further, we consider the system of $N$ identical particles on the torus
$\bT$. We shall assume that the Hamiltonian for $N$ bosons or $N$ fermions
has the form:
\begin{equation}\label{bol6}
\widehat{H}_N=-\frac{\hbar^2}{2m}\sum^{N}_{j=1}\Delta_j
+\sum^{N}_{j=1}\sum^{N}_{l=j+1}V(x_j-x_l).
\end{equation}
According to  \cite{Book_Ultravt}, in the boson case, to this operator corresponds
the ultra secondary quantized Hamiltonian \begin{eqnarray}\label{bol7}
&&\ol{\widehat{H}}_B
=\sum^{\infty}_{k=0}\sum^{\infty}_{M=0}\frac1{k!M!}
\sum^{\infty}_{s_1=0}\dots\sum^{\infty}_{s_k=0}\int\dots\int
dx_1\dots dx_kdy_1\dots dy_{2M}\times
\nn\\
&&\quad \times \widehat{b}^+(x_1,s_1)\cdot\dots\cdot
\widehat{b}^+(x_k,s_k) \widehat{B}^+(y_1,y_2)\cdot\dots\cdot
\widehat{B}^+(y_{2M-1},y_{2M}) \widehat{H}_{k+2M}\times
\nn\\
&&\quad \times \Symm_{x_1\dots x_ky_1\dots y_{2M}}
\Big(\widehat{b}^-(x_1,s_1)\cdot\dots\cdot \widehat{b}^-(x_k,s_k)
\widehat{B}^-(y_1,y_2)\cdot\dots
\nn\\
&&\qquad\qquad\qquad\qquad\qquad\qquad \cdot
\widehat{B}^-(y_{2M-1},y_{2M})\Big)\times
\\
&&\quad \times \exp\bigg(-\sum^{\infty}_{s=0}\int dx
\widehat{b}^+(x,s)\widehat{b}^-(x,s) -\iint dy dy'
\widehat{B}^+(y,y')\widehat{B}^-(y,y')\bigg), \nn ,
\end{eqnarray}
while in the fermion case the corresponding operator is
$\ol{\widehat{H}}_F$ and is expressed by a similar formula in which
$\Symm$ is replaced by  $\Asymm$. As in  (\ref{bol6}) and
(\ref{bol7}), the ultrasecondary quantized operator
$\ol{\widehat{A}}$ is assigned  \cite{Book_Ultravt} to any
$N$-particle operator
$$
\widehat{A}_N\bigg(\str{2}{x_1},\dots,\str{2}{x_N};
-i\str{1}{\frac{\pa}{\pa x_1}},\dots, -i\str{1}{\frac{\pa}{\pa
x_N}}\bigg).
$$

For example, to the unit operator in the boson case, we assign the following
ultrasecondary quantized unit operator:
\begin{eqnarray}\label{bol8}
&&\ol{\widehat{E}}_B
=\sum^{\infty}_{k=0}\sum^{\infty}_{M=0}\frac1{k!M!}
\sum^{\infty}_{s_1=0}\dots\sum^{\infty}_{s_k=0}\int\dots\int
dx_1\dots dx_kdy_1\dots dy_{2M}\times
\nn\\
&&\quad \times \widehat{b}^+(x_1,s_1)\cdot\dots\cdot
\widehat{b}^+(x_k,s_k) \widehat{B}^+(y_1,y_2)\cdot\dots\cdot
\widehat{B}^+(y_{2M-1},y_{2M})\times
\nn\\
&&\quad \times \Symm_{x_1\dots x_ky_1\dots y_{2M}}
\Big(\widehat{b}^-(x_1,s_1)\cdot\dots\cdot \widehat{b}^-(x_k,s_k)
\widehat{B}^-(y_1,y_2)\cdot\dots
\nn\\
&&\qquad\qquad\qquad\qquad\qquad\qquad \cdot
\widehat{B}^-(y_{2M-1},y_{2M})\Big) \times\nn \\
&&\quad \times\exp\bigg(-\sum^{\infty}_{s=0}\int dx
\widehat{b}^+(x,s)\widehat{b}^-(x,s) -\iint dy dy'
\widehat{B}^+(y,y')\widehat{B}^-(y,y')\bigg), \nn ,
\end{eqnarray}
which is the sum of the projectors ~(\ref{bol4}).
Similarly, in the fermion case, the ultrasecondary quantized unit operator is
$$
\ol{\widehat{E}}_F=\sum^{\infty}_{k=0}\sum^{\infty}_{M=0}
\widehat{\Pi}^{\rm Asymm}_{k,M},
$$
with $\Symm$ replaced by $\Asymm$ in formula ~(\ref{bol8}).

Consider the following eigenvalue problem

\begin{equation}\label{bol9}
\ol{\widehat{H}}_{B,F}\Phi=\lambda\ol{\widehat{E}}_B\Phi, \qquad
\ol{\widehat{E}}\Phi\ne0,
\end{equation}
in the Bose and Fermi cases.
The following assertion, proved in
\cite{Book_Ultravt} is valid:
{\textit{in the subspaces  $\cF^{\rm Symm}_{k,M}$
and $\cF^{\rm Asymm}_{k,M}$ of the space  $\cF$,
the operators $\ol{\widehat{H}}_B$ and $\ol{\widehat{H}}_F$
coincide  with the operator $\widehat{H}_{k+2M}$}}.
Therefore, the eigenvalues $\lambda$ of  problem ~(\ref{bol9}) in
the boson and fermion cases coincide with the corresponding
eigenvalues of the operators  $\widehat{H}_N$~(\ref{bol6}). For the
case in which the commutators between the operators  $\widehat{b}^-(x,s)$ and
$\widehat{b}^+(x,s)$, as well as the operators $\widehat{B}^-(x,y)$ and
$\widehat{B}^+(x,y)$ are $1/N$ small, the asymptotics of the solutions of the problem
(\ref{bol9}), according to  \cite{Book_Ultravt}, are determined by the extremum points
of the symbol corresponding to problem ~(\ref{bol9}). In the boson case,
the symbol has the form
\begin{eqnarray}\label{bol10}
&&\cH_B[b^*(\cdot),b(\cdot),B^*(\cdot),B(\cdot)]
\nn\\
&&\quad =\bigg\{\sum^{\infty}_{k,M=0}\frac1{k!M!}
\sum^{\infty}_{s_1=0}\dots\sum^{\infty}_{s_k=0} \int\dots\int
dx_1\dots dx_k dy_1\dots dy_{2M}\times
\nn\\
&&\quad \times b^*(x_1,s_1)\cdot\dots\cdot b^*(x_k,s_k)
B^*(y_1,y_2)\cdot\dots\cdot B^*(y_{2M-1},y_{2M})H_{k+2M}\times
\nn\\
&&\quad \times \Symm_{x_1\dots x_k y_1\dots y_{2M}}
\Big(b(x_1,s_1)\cdot\dots\cdot b(x_k,s_k)
B(y_1,y_2)\cdot\dots\cdot B(y_{2M-1},y_{2M})\Big)\bigg\}\times
\nn\\
&&\quad \times \bigg\{\sum^{\infty}_{k',M'=0}\frac1{k'!M'!}
\sum^{\infty}_{s'_1=0}\dots\sum^{\infty}_{s'_{k'}=0}
\int\dots\int dx'_1\dots dx'_{k'} dy'_1\dots dy'_{2M'}\times
\nn\\
&&\quad \times b^*(x'_1,s'_1)\cdot\dots\cdot b^*(x'_{k'},s'_{k'})
B^*(y'_1,y'_2)\cdot\dots\cdot B^*(y'_{2M'-1},y'_{2M'})\times
\\
&&\quad\times \Symm_{x'_1\dots x'_{k'} y'_1\dots y'_{2M'}}
\Big(b(x'_1,s'_1)\cdot\dots\cdot b(x'_{k'},s'_{k'})
B(y'_1,y'_2)\cdot\dots\cdot B(y'_{2M'-1},y'_{2M'})\Big) \bigg\}.
\nn
\end{eqnarray}

In the Fermi case, the symbol is expressed similarly, except that  $\Symm$
in formula (\ref{bol10}) is replaced by $\Asymm$.

The following identity for the symbol   (\ref{bol10}) holds in the Bose case:

\begin{equation}\label{bol11}
\cH_B[b^*(\cdot),b(\cdot),B^*(\cdot),B(\cdot)]
=\frac{\Sp(\widehat\rho_B\widehat{H})}{\Sp(\widehat\rho_B)},
\end{equation}
where $\widehat{H}$, $\widehat{\rho}_B$ are the secondary quantized operators
\begin{eqnarray}\label{bol12}
\widehat{H}&=&\int dx\widehat{\psi}^+(x)
\bigg(-\frac{\hbar^2}{2m}\Delta\bigg)\widehat{\psi}^-(x)+
\nn\\
&&\qquad +\frac12\int\int dx dy V(x,y)
\widehat{\psi}^+(y)\widehat{\psi}^+(x)
\widehat{\psi}^-(y)\widehat{\psi}^-(x).
\end{eqnarray}
Here $\widehat{\rho}_B$ depends on the functions $b(x,s)$, $B(y,y')$:
\begin{eqnarray}\label{bol13}
\widehat{\rho}_B&=&\sum^{\infty}_{k=0}\sum^{\infty}_{M=0}
\frac1{k!M!(k+2M)!} \bigg(\sum^{\infty}_{s=0}\int\int dx dx'
b(x,s)b^*(x',s)
\widehat{\psi}^+(x)\widehat{\psi}^-(x')\bigg)^k\times
\nn\\
&&\quad \times \bigg(\iint dy_1 dy_2 B(y_1,y_2)
\widehat{\psi}^+(y_1)\widehat{\psi}^+(y_2)\bigg)^M\times
\nn\\
&&\quad \times \bigg(\iint dy'_1 dy'_2 B(y'_1,y'_2)
\widehat{\psi}^-(y'_1)\widehat{\psi}^-(y'_2)\bigg)^M\times
\nn\\
&&\quad \times \exp\bigg(-\int dz
\widehat{\psi}^+(z)\widehat{\psi}^-(z)\bigg),
\end{eqnarray}
where $\widehat{\psi}^+(x)$, $\widehat{\psi}^-(x)$
are the Bose creation and annihilation operators, ordered according top Vick ~\cite{7}.
In the Fermi case we have a similar identity
$$
\cH_F[b^*(\cdot),b(\cdot),B^*(\cdot),B(\cdot)]
=\frac{\Sp(\widehat\rho_F\widehat{H})}{\Sp(\widehat\rho_F)},
$$
where $\widehat{H}$, $\widehat{\rho}_F$ are the following secondary quantized
operators:
$$
\widehat{H}=\int dx\widehat{\psi}^+(x)
\bigg(-\frac{\hbar^2}{2m}\Delta\bigg)\widehat{\psi}^-(x) \qquad
\qquad \qquad \qquad \qquad \qquad
$$
$$
\qquad \qquad \qquad \qquad +\frac12\int\int dx dy V(x,y)
\widehat{\psi}^+(x)\widehat{\psi}^+(y)
\widehat{\psi}^-(y)\widehat{\psi}^-(x)
$$
and
\begin{eqnarray}\label{bol14}
\widehat{\rho}_F&=&\sum^{\infty}_{k=0}\sum^{\infty}_{M=0}
\frac1{k!M!(k+2M)!} \bigg(\iint dy_1 dy_2 B(y_1,y_2)
\widehat{\psi}^+(y_1)\widehat{\psi}^+(y_2)\bigg)^M \times
\nn\\
&&\quad \times \sum^{\infty}_{s_1=0}\dots\sum^{\infty}_{s_k=0}
\int\dots\int dx_1 dx'_1\dots dx_k dx'_k\times
\nn\\
&&\quad \times b(x_1,s_1)b^*(x'_1,s_1)\cdot\dots\cdot
b(x_k,s_k)b^*(x'_k,s_k)\times
\nn\\
&&\quad \times
\widehat{\psi}^+(x_1)\cdot\dots\cdot\widehat{\psi}^+(x_k)
\widehat{P}_0
\widehat{\psi}^-(x'_k)\cdot\dots\cdot\widehat{\psi}^-(x'_1)\times
\nn\\
&&\quad \times \bigg(\iint dy'_1 dy'_2 B(y'_1,y'_2)
\widehat{\psi}^-(y'_2)\widehat{\psi}^-(y'_1)\bigg)^M;
\end{eqnarray}
here, in the given case,  $\widehat{\psi}^+(x)$, $\widehat{\psi}^-(x)$
are the Fermi creation and annihilation operators and $\widehat{P}_0$
is the projector on the vacuum vector of the fermionic Fock space.
In general, for an arbitrary secondary quantized operator $\widehat{A}$,
the symbol of the corresponding ultrasecondary quantized operator
$\ol{\widehat{A}}$ is expressed
\cite{Book_Ultravt} by the following formula:
$$
A_{B,F}[b^*(\cdot),b(\cdot),B^*(\cdot),B(\cdot)]
=\frac{\Sp(\widehat\rho_{B,F}\widehat{A})}{\Sp(\widehat\rho_{B,F})}.
$$
In the space $\cF$, we introduce  \cite{Book_Ultravt} the ultrasecondary quantized operators for the number of particles:
\begin{equation}\label{bol15}
\ol{\widehat{N}}_B=\sum^{\infty}_{k=0}\sum^{\infty}_{M=0}(k+2M)
\widehat{\Pi}^{\rm Symm}_{k,M}, \qquad
\ol{\widehat{N}}_F=\sum^{\infty}_{k=0}\sum^{\infty}_{M=0}(k+2M)
\widehat{\Pi}^{\rm Asymm}_{k,M}.
\end{equation}
Correspondingly, in the boson case, the operator
$\ol{\widehat{N}}_B$ has the form
\begin{eqnarray}\label{bol16}
&&N_B=\bigg\{\sum^{\infty}_{k=0}\sum^{\infty}_{M=0}
\frac{k+2M}{k!M!} \sum^{\infty}_{s_1=0}\dots\sum^{\infty}_{s_k=0}
\int\dots\int dx_1 \dots dx_{k+2M}\times
\nn\\
&&\quad \times b^*(x_1,s_1)\cdot\dots\cdot b^*(x_k,s_k)
B^*(x_{k+1},x_{k+2})\cdot\dots\cdot B^*(x_{k+2M-1},x_{k+2M})\times
\nn\\
&&\quad \times \Symm_{x_1\dots x_{k+2M}}
\Big(b(x_1,s_1)\cdot\dots\cdot b(x_k,s_k)
B(x_{k+1},x_{k+2})\cdot\dots
\nn\\
&&\qquad\qquad \qquad\qquad \qquad\qquad \cdot
B(x_{k+2M-1},x_{k+2M})\Big)\bigg\}\times
\nn\\
&&\quad \times \bigg\{\sum^{\infty}_{k'=0}\sum^{\infty}_{M'=0}
\frac{1}{k'!M'!}
\sum^{\infty}_{s'_1=0}\dots\sum^{\infty}_{s'_{k'}=0}
\int\dots\int dz_1 \dots dz_{k'+2M'}
\nn\\
&&\quad \cdot b^*(z_1,s'_1)\cdot\dots\cdot b^*(z_{k'},s'_{k'})
B^*(z_{k'+1},z_{k'+2})\cdot\dots\cdot
B^*(x_{k'+2M'-1},x_{k'+2M'})\times
\nn\\
&&\quad \times \Symm_{z_1\dots z_{k'+2M'}}
\Big(b(z_1,s'_1)\cdot\dots\cdot
B(z_{k'+2M'-1},z_{k'+2M'})\Big)\bigg\}^{-1}.
\end{eqnarray}
In the corresponding formula for the fermion case,
 $\Symm$ is replaced by  $\Asymm$.

\section{Symbol of the ultrasecondary quantized operator}

First of all let us note that the definition of the symbol given above does not reflect the
the thermodynamical asymptotics, although this definition is in accord with the
Bogolyubov--Dirak rule asserting that the creation and annihilation operators in the leading term of the asymptotics must be replaced by $c$-numbers. However, we shall say that the symbol obtained in this way is a {\it pseudosymbol}. Suppose that the operator
 $\widehat{H}$ is of the form
\begin{eqnarray}\label{ultra1}
&&\widehat{H}=\sum_{l=1}^{L} \int \dots\int dx_1 \dots
dx_l\widehat{\psi}^+(x_1)\dots \widehat{\psi}^+(x_l) \times \nn \\
&& \times H_l \bigg(\str{2}{x_1},\dots,\str{2}{x_l}; -i
\frac{\str{1}{\pa}}{\pa x_1}, \dots, -i\frac{\str{1}{\pa}}{\pa
x_l}\bigg) \widehat{\psi}^-(x_l) \dots \widehat{\psi}^-(x_1).
\end{eqnarray}
Then, in the case of ultrasecondary quantization without the creation and annihilation
operators of pairs of particles  $\widehat{B}^{\pm}(x,y)$ for the operators
$\ov{\widehat{H}}$ and  $\ov{\widehat{E}}$ defined above, we have the identity
\begin{equation}\label{ultra2}
\ov{\widehat{H}}=\ov{\widehat{E}} \ov{\widehat{A}},
\end{equation}
where $\ov{\widehat{A}}$ is the operator on the space  $\cF$,
of the form
\begin{eqnarray}\label{ultra3}
&&\ov{\widehat{A}} =\sum_{l=1}^{L}\sum_{s_1=0}^\infty \dots
\sum_{s_l=0}^\infty\int \dots \int
 dx_1 \dots dx_l\widehat{b}^+(x_1,s_1)\dots \widehat{\psi}^+(x_l,s_l) \times \nn \\
&& \times H_l \bigg(\str{2}{x_1},\dots,\str{2}{x_l}; -i
\frac{\str{1}{\pa}}{\pa x_1}, \dots, -i\frac{\str{1}{\pa}}{\pa
x_l}\bigg) \widehat{b}^-(x_l,s_l) \dots \widehat{b}^-(x_1).
\end{eqnarray}

For the case in which the ultrasecondary quantization also takes into account
the creation and annihilation operators for pairs of particles, the identity
(\ref{ultra2}) remains valid, except that the operator  $\ov{\widehat{A}}$
has a more complicated expression than the one in (\ref{ultra3}).
For example, when we consider the operator  $\widehat{H}$
in the particular case of the formula  (\ref{ultra1}) with
$L=2$, the operator $\ov{\widehat{A}}$  will have the form
\begin{eqnarray}\label{ultra4}
&&\ov{\widehat{A}}=\sum_{s=0}^\infty \int
dx\widehat{b}^+(x,s)\bigg(-\frac{h^2}{2m}\Delta\bigg)
\widehat{b}^-(x,s)+ \nn \\
&& +\iint dx \ dy \widehat{B}^+(x,y)
\bigg(-\frac{h^2}{2m}(\Delta_x+\Delta_y)\bigg)\widehat{B}^-(x,y)+
\nn \\
&&+ \frac 12 \sum_{s_1=0}^\infty \sum_{s_2=0}^\infty \iint dx \ dy
V(x,y)\widehat{b}^+(x,s_1)\widehat{b}^+(y,s_2)\widehat{b}^-(y,s_2)\widehat{b}^-(x,s_1)+\nn\\
&&+\sum_{s=0}^\infty \iiint dx \ dy \ dz(V(x,y)+V(x,z)) \times
\widehat{b}^+(x,s)\widehat{B}^+(y,z)\widehat{B}^-(y,z)\widehat{b}^-(x,s)+\nn\\
&&+\iint dx \ dyV(x,y)\widehat{B}^+(x,y)\widehat{B}^-(x,y)+\frac
12 \iiiint dx \ dy \ dz\ dw
V(x,y)\widehat{B}^+(x,y)\widehat{B}^+(z,w) \times \nn \\
&&\times \bigg(\widehat{B}^-(y,w)\widehat{B}^-(x,z)+
\widehat{B}^-(w,y)\widehat{B}^-(z,x)+\widehat{B}^-(y,z)\widehat{B}^-(w,x)+\widehat{B}^-(z,y)\widehat{B}^-(x,w)
\bigg).
\end{eqnarray}
If, in the expressions (\ref{ultra3}), (\ref{ultra4}), we replace the operators
 $\widehat{B}^\pm(x,y), \widehat{b}^\pm(x,y)$ by
$c$-numbers, we obtain the symbol corresponding to the asymptotics in the thermodynamical limit.

Consider the system of $N$ identical bosons of mass $m$ locate in the three-dimensional rectangle $\bT$ with side lengths
$L_1$, $L_2$, and $L_2$.
We assume that the bosons interact with interaction potential

\begin{equation}
V\left(N^{1/3}(x-y)\right), \label{1aa}
\end{equation}
where $V(\xi)$ is a finite even function,  $x,y$ are the coordinates of the boson
in the rectangle $\bT$. The boundary conditions along $L_1$
will be assumed periodic, while along $L_2$ we impose the condition of having zero derivatives. Note that the potential ~(\ref{1aa})
depends on $N$ in such a way that the radius of the interaction potential
decreases with the increase of the number of particles $N$, but so that
on the average the number of particles which interact with a fixed particle remains constant.

Under ultrasecondary quantization for pairs of bosons in the considered system,
we obtain the (ultrasecondary quantized) operator
$\ov{\wh{H}}$ whose explicit form was presented above.
As was explained previously, for this ultrasecondary quantized operator we have the identity
\begin{equation}
\ov{\wh{H}}=\ov{\wh{E}}\wh{A}, \label{2aa}
\end{equation}
where  $\ov{\wh{E}}$ is the ultrasecondary quantized unit operator, while $\wh{A}$
is an operator in the space of ultrasecondary quantization. It is easy to verify that
the following operator satisfies an identity of the form~(\ref{2aa}):

\begin{eqnarray}
&&\wh{A}=\iint
dxdy\wh{B}^+(x,y)\left(-\frac{\hbar^2}{2m}(\Delta_x+\Delta_y)+
V\left(N^{1/3}(x-y)\right)\right)\wh{B}^-(x,y)+\nn\\
&&+2\int dxdydx'dy'
V\left(N^{1/3}(x-y)\right)\wh{B}^+(x,y)\wh{B}^+(x',y')
\wh{B}^-(x,x')\wh{B}^-(y,y'), \label{3aa}
\end{eqnarray}
where $\wh{B}^+(x,y)$ and $\wh{B}^-(x,y)$ are respectively the boson creation and annihilation operators for pairs of particles in the Fock space of ultrasecondary quantization. By the identity ~(\ref{2aa}), in order to find the asymptotics of the spectrum of the boson system under consideration in the limit as $N\to\infty$
we must find the corresponding asymptotics for the operator
~(\ref{3aa}).

Since the function~(\ref{1aa}) of multiplication by $N$ in the limit as
$N\to\infty$ converges in the weak sense to the Dirak delta function, it follows that
the second summand in the operator ~(\ref{3aa}) in this limiting case contains the
small parameter $1/N$. This means that, in order to find the asymptotics of the eigenvalues and the eigenfunctions of the operator  $\wh{A}$
one can apply the quasiclassical methods developed in \cite{Masl_Shved}.
The asymptotics of the eigenvalues and the eigenfunctions is determined by
the symbol of the operator ~(\ref{3aa}); this symbol is called the {\it true symbol}
of the ultrasecondary quantized problem.
The true symbol corresponding to the operator
~(\ref{3aa}) is the following functional, defined for any pair of functions
 $\Phi^+(x,y)$ and $\Phi(x,y)$:

\begin{eqnarray}
&&\cH\left[\Phi^+(\cdot),\Phi(\cdot)\right]=\iint
dxdy\Phi^+(x,y)\left(
-\frac{\hbar^2}{2m}(\Delta_x+\Delta_y)\right)\Phi(x,y)+\nn\\
&&+2\int dxdydx'dy'
\left(NV\left(N^{1/3}(x-y)\right)\right)\Phi^+(x,y)
\Phi^+(x',y')\Phi(x,x')\Phi(y,y'). \label{4aa}
\end{eqnarray}
From the invariance of the number of particles in the system with the functions
 $\Phi^+(x,y)$ and $\Phi(x,y)$, we obtain the condition

\begin{equation}
\iint dxdy \Phi^+(x,y)\Phi(x,y)=\frac12. \label{5aa}
\end{equation}

According to the asymptotic methods (see \cite{Masl_Shved}),
to each solution of the system  of equations

\begin{equation}\label{6aa}
\Omega\Phi(x,y)=\frac{\delta\cH}{\delta\Phi^+(x,y)},\qquad
\Omega\Phi^+(x,y)=\frac{\delta\cH}{\delta\Phi(x,y)}
\end{equation}
which also satisfies condition ~(\ref{5aa}), there corresponds, in the limit as
$N\to\infty$, the asymptotic series of eigenvalues and eigenfunctions of the operator
~(\ref{3aa}). From the expression for the true symbol
~(\ref{4aa}), it follows that the system of equations ~(\ref{6aa})
can be written in the form

\begin{equation}\label{7aa}
\begin{aligned}
&\Omega\Phi(x,y)=-\frac{\hbar^2}{2m}(\Delta_x+\Delta_y)\Phi(x,y)+\nn\\
&+\iint
dx'dy'\left(NV\left(N^{1/3}(x-y)\right)+NV\left(N^{1/3}(x'-y')\right)
\right)\Phi^+(x',y')\Phi(x,x')\Phi(y,y'),\nn\\
&\Omega\Phi^+(x,y)=-\frac{\hbar^2}{2m}(\Delta_x+\Delta_y)\Phi^+(x,y)+\nn\\
&+\iint
dx'dy'\left(NV\left(N^{1/3}(x-x')\right)+NV\left(N^{1/3}(y-y')\right)
\right)\Phi(x',y')\Phi^+(x,x')\Phi^+(y,y').
\end{aligned}
\end{equation}

Let  $v_q$ be the Fourier coefficients on the box $(L_1,L_2,L_2)$ of the potential  $NV(N^{1/3} x)$:

\begin{equation}\label{7aa1}
v_q=\frac{1}{L_1L_2^2} \int_T
e^{-iqx}NV\bigg(\sqrt[3]{N}(x)\bigg)dx.
\end{equation}

The exact solution of system
~(\ref{7aa}) is given by the following functions:

\begin{eqnarray}
&&\Phi^+_{k_1,k_2}= \frac{1}{L_1L_2^2}  e^{-ik_1(x+y)}
\cos\big(k_2(x-y)\big); \label{7aa2}\\  &&\Phi_{k_1,k_2}=
\frac{1}{L_1L_2^2} \sum \varphi_{k_2,l} e^{ik_1(x+y)} e^{il(x-y)}
\label{7aa2'}
\end{eqnarray}
with eigenvalues

\begin{equation}\label{7aa3}
\Omega =\frac{h^2}{m} (k_1^2+k_2^2)+v_0+v_{2k_2},
\end{equation}
where the function  $\varphi_{k_2,l}$ is given by
\begin{eqnarray}\label{7aa4}
&&\varphi_{k_2,l} = -\frac{b_l}{2}+\frac 12\sqrt{b^2_l -1}, \qquad
l^2>k_2^2; \nn \\
&&\varphi_{k_2,l} = -\frac{b_l}{2}-\frac 12\sqrt{b^2_l -1}, \qquad
l^2<k_2^2; \nn \\
&&b_l=\frac{h^2/m(l^2-k^2_2)-(v_0+v_{2k_2})}{v_{l-k_2}+v_{l+k_2}},
\qquad \varphi_{k_2,k_2}=\frac 12. \nn
\end{eqnarray}

The pair of vectors $k_1$, $k_2$ plays the role of parameters indexing the various solutions of this system. The vector $\hbar k_1/m$
expresses the velocity of the flow of the boson system along the capillary. The vector
 $k_2$ is the wave vector of the transversal mode.

Note that as  $|l| \to\infty$ we have  $b_l \to \infty$, since

\begin{equation}\label{7aa4''}
|v_l|=\frac{1}{L_1L_2^2} \int _{NT} e^{-il\xi/N}V(\xi)d\xi\leq
\frac{1}{L_1L_2^2}\int _{NT}\bigl| V(\xi)\bigr|d\xi<
\frac{1}{L_1L_2^2}\int_{R^3}V(\xi) d\xi\leq v_0,
\end{equation}

so that the series (\ref{7aa2'}) converges absolutely.

Let us split the series (\ref{7aa2'}) into two parts for $l\leq N^{1/6}$ and for
$l> N^{1/6}$. The first sum converges as $N \to\infty$ with precision up to
$(N^{-1/6})$ to the value

\begin{eqnarray}\label{7aa5}
&&b_l =\frac{h^2(l^2-k^2_2)}{2mV_0} -1, \nn \\
&& \varphi_{k_2,l} = - \frac{b_l}{2}\pm \frac 12\sqrt{b^2_l -1}.
\end{eqnarray}
This follows easily from the substitution $\sqrt[3]{N}x=\xi$ in equation
 (\ref{7aa1}).

The second part of the sum tends to zero by (\ref{7aa4'}) up to
$O(N^{-1/6}).$ Therefore in the limit as  $N\to\infty$, the system of equations
~(\ref{7aa}) under the additional condition ~(\ref{5aa}) possesses, for
$k_1=0$, the following family of solutions:

\begin{eqnarray}
&&\Phi^+_{k}(x,y)=\frac1{L_1L_2^2}\cos\left(k(x-y)\right),\nn\\
&&\Phi_{k}(x,y)=\frac1{L_1L_2^2}\sum_{l}\varphi_{k,l}\exp\left(il(x-y)\right),
\label{8aa}
\end{eqnarray}
where $k$ and $l$ are three-dimensional vectors of the form
$$
2\pi\left(0,\frac{n_2}{L_2},\frac{n_3}{L_2}\right);
$$
here $n_2$, $n_3$ are integers and the terms $\varphi_{k,l}$ in formula ~(\ref{2aa})
have the form

\begin{equation}
\varphi_{k,l}=\frac1{2V_0}\left(\frac{\hbar^2}{2m}(k^2-l^2)+V_0\pm
\sqrt{\left(\frac{\hbar^2}{2m}(k^2-l^2)+V_0\right)^2-V_0^2}\right),
\label{9aa}
\end{equation}
where the plus sign is chosen when
$l^2>k^2$, the minus sign, when
$l^2<k^2$, and, finally,  $V_0$ here, as before, stands for the quantity

\begin{equation}
V_0=\frac1{L_1L_2^2}\int dx V(x), \label{10aa}
\end{equation}
in which the integral is taken over the space  $\bR^3$.

The vector $k$ in formulas (\ref{8aa}) plays the role of a parameter indexing
the various solutions of equations \eqref{7aa}, \eqref{5aa}.
The solutions \eqref{8aa} are standing waves and correspond to series
for which there is no flow.

The leading term of the asymptotics of the eigenvalues of the series corresponding
to the solution of \eqref{7aa2} \eqref{7aa2'} is equal to the value
of the symbol  \eqref{4aa} on these functions multiplied by $N$:
\begin{equation}\label{11aa}
E_{k_1,k_2}=N\l\frac{\hbar^2(k_1^2+k_2^2)}{2m}+\frac{V_{0}}{2}\r.
\end{equation}

The asymptotics of the eigenvalues and eigenfunctions, in particular the terms
that follow  $E_{k_1,k_2}$, are determined not only by the system \eqref{7aa},
but also by the solution of the variational system for the Hamiltonian system
of equations. The system of variational equations for
\eqref{7aa} has the form:

\begin{equation}\label{Variational Equation System for any N}
\begin{aligned}
    &\l\Omega-\lambda\r F(x,y)=-\frac{\hbar^2}{2m}(\Delta_x+\Delta_y)F(x,y)+\\&+
    2N\iint\,dx'\,dy'\l V\big(\sqrt[3]{N}(x-y)\big)+V\big(
    \sqrt[3]{N}(x'-y')\big)\r\big( G(x',y')\Phi(x,x')\Phi(y,y')+\\&+
    \Phi^{+}(x',y')F(x,x')\Phi(y,y')+
    \Phi^{+}(x',y')\Phi(x,x')F(y,y')\big),\\
    &\l\Omega+\lambda\r G(x,y)=-\frac{\hbar^2}{2m}(\Delta_x+\Delta_y)G(x,y)+\\&+
    2N\iint\,dx'\,dy'\l V\big(\sqrt[3]{N}(x-x')\big)+V\big(
    \sqrt[3]{N}(y-y')\big)\r\big( F(x',y')\Phi^{+}(x,x')\Phi^{+}(y,y')+\\&+
    \Phi(x',y')G(x,x')\Phi^{+}(y,y')+
    \Phi(x',y')\Phi^{+}(x,x')G(y,y')\big).
\end{aligned}
\end{equation}
In order to find the spectrum of the quasiparticles, one must distinguish,
among all the solutions of the variational system, those which satisfy the
selection rule
\[
\iint dx\,dy\,\l G^{*}(x,y)G(x,y)-F^{*}(x,y)F(x,y)\r>0.
\]

If $k_2=0$, then the asymptotic series corresponding to this solution is the Bogolyubov series with flow velocity  $\hbar k_1/m$.
The spectrum of quasiparticles of this series is expressed by the well known
formula

\begin{equation}\label{Bogolubov Spectrum}
\lambda_l=\sqrt{\left(\frac{\hbar^2l^2}{2m}+V_{0}^{2}\right)-V_{0}^2}
-\frac{\hbar^2lk_1}{m}.
\end{equation}

Consider the case  $k_2\neq 0$. Substituting the solution \eqref{7aa2},
\eqref{7aa2'} into \eqref{Variational Equation System for any N} and taking
symmetry into account, we obtain the following expression for the
solution of the variational system:

\begin{eqnarray}\label{Variational Maslov's decision}
&&G_{l}(x,y)=\frac{u_{1,l}}{2}\Big( \exp(
i(k_1+k_2)x+i(k_1+l)y)+\exp(
    i(k_1+k_2)y+i(k_1+l)x)\Big)+ \nn \\
    &&+\frac{u_{2,l}}{2}\Big( \exp( i(k_1-k_2)x+i(k_1+2k_2+l)y)+ \nn \\
    &&+\exp(i(k_1-k_2)y+i(k_1+2k_2+l)x)\Big),\\
    &&F_{l}(x,y)=\frac{v_{1,l}}{2}\Big( \exp(
    i(k_1+k_2)x+i(k_1+l)y)+\exp(
    i(k_1+k_2)y+i(k_1+l)x)\Big)+\nn \\
    &&+\frac{v_{2,l}}{2}\Big( \exp( i(k_1-k_2)x+i(k_1+2k_2+l)y)+ \nn \\
    &&+\exp(
    i(k_1-k_2)y+i(k_1+2k_2+l)x)\Big)+ \nn \\
    &&+\sum_{l'\neq l,l+2k_2}\frac{w_{l,l'}}{2}\Big(\exp( i(k_1+k_2+l-l')x+i(k_1+l')y)+ \nn \\
    &&+\exp\l i(k_1+k_2+l-l')y+i(k_1+l')x\r\Big), \nn
\end{eqnarray}
where  $l\neq-k_2$, while the numerical coefficients  $u_{1,l}$, $u_{2,l}$,
$v_{1,l}$, $v_{2,l}$, $w_{l,l'}$ are determined by an infinite system of equations. This system contains a closed subsystem consisting of four
equations for the coefficients  $u_{1,l}$,
$u_{2,l}$, $v_{1,l}$, $v_{2,l}$, which can be written in standard form
 \begin{equation}\label{fortildelabbda}
\wt{\lambda} X=MX,
\end{equation}
where
$$
\wt{\lambda}=\lambda-\frac{\hbar^2}{m}k_1(k_2+l);
$$
here $X$ is the column vector
$$
X=\left(
\begin{array}{c}
u_{1,l}\\
u_{2,l}\\
v_{1,l}\\
v_{2,l}
\end{array}
\right),
$$
$M$ is the matrix
$$
M=\left(
\begin{array}{cccc}
B_{l} & V_{0} & -V_{0} & 0\\
V_{0} & B_{l+2k_2} & 0 & -V_{0}\\
2V_{0}\varphi_{k_2,l} & V_{0}\l\varphi_{k_2,l}+\varphi_{k_2,l+2k_2}\r & -B_{l} & -V_{0}\\
V_{0}\l\varphi_{k_2,l}+\varphi_{k_2,l+2k_2}\r &
2V_{0}\varphi_{k_2,l+2k_2} & -V_{0} & -B_{l+2k_2}
\end{array}
\right),
$$
and the $B_{l}$ have the form
$$
B_{l}=\frac{\hbar^2}{2m}(l^2-k_2^2)+V_{0}\varphi_{k_2,l}.
$$

The matrix $M$ can be written in block form

$$
M=\left(
\begin{array}{cc}
C & -V_{0}E \\
D & -C
\end{array}
\right),
$$
constituted by the following  $2\times2$ matrices:
$$
C=\left(
\begin{array}{cc}
B_{l} & V_{0} \\
V_{0} & B_{l+2k_2}
\end{array}
\right),\quad D=\left(
\begin{array}{cc}
2V_{0}\varphi_{k_2,l} &
V_{0}\l\varphi_{k_2,l}+\varphi_{k_2,l+2k_2}\r\\
V_{0}\l\varphi_{k_2,l}+\varphi_{k_2,l+2k_2}\r &
2V_{0}\varphi_{k_2,l+2k_2}
\end{array}
\right)
$$
and by the unit matrix $E$.

This expression allows us to rewrite equation (\ref{fortildelabbda})
$$
\wt{\lambda}^2u=(C^2-V_0D)u
$$
for the column vector $u=\left(
\begin{array}{c}
u_{1,l}\\
u_{2,l}\\
\end{array}
\right)$.

The selection rule  (\ref{Variational Maslov's decision}) for the
variational system of equations, which has the form
$$
u_{1,l}^{*}u_{1,l}+u_{2,l}^{*}u_{2,l}-v_{1,l}^{*}v_{1,l}-v_{2,l}^{*}v_{2,l}>0.
$$
specifies the following eigenvalues:
\begin{equation}\label{lambda 1 2 k1 k2 l}
\begin{aligned}
    &\lambda_{1,k_1,k_2,l}=-\frac{\hbar^2}{m}k_1(k_2+l)\pm\\&\pm\frac{\hbar^2}{2m}
    \sqrt{k_2^4+\frac{l^4}{2}+\frac{l_1^4}{2}-k_2^2(l^2+l_1^2)+\frac{1}{2}
    (l^2+l_1^2-2k_2^2)\sqrt{(l^2-l_1^2)+(\frac{4mV_{0}}{\hbar^2})}},\\
    &\lambda_{2,k_1,k_2,l}=-\frac{\hbar^2}{m}k_1(k_2+l)\pm\\&\pm\frac{\hbar^2}{2m}
    \sqrt{k_2^4+\frac{l^4}{2}+\frac{l_1^4}{2}-k_2^2(l^2+l_1^2)-\frac{1}{2}
    (l^2+l_1^2-2k_2^2)\sqrt{(l^2-l_1^2)+(\frac{4mV_{0}}{\hbar^2})}},
\end{aligned}
\end{equation}
where $l_1=l+2k_2$ and $l\neq k_2$. In the formula for
$\lambda_{1,k_1,k_2,l}$, we choose the plus sign if $l^2>k_2^2$, and
the minus sign if  $l^2<k_2^2$, whereas  in the formula for
$\lambda_{2,k_1,k_2,l}$, we take the plus sign if $l_1^2>k_2^2$, and minus,
if $l_1^2<k_2^2$.

Formulas \eqref{lambda 1 2 k1 k2 l} for $l\neq -k_2$ determine the spectrum
of the quasiparticles of the series corresponding to the solutions of
\eqref{7aa2}, \eqref{7aa2'}. From the explicit form  (\ref{lambda 1 2 k1 k2 l}) it
follows that there are negative values in the spectrum of the quasiparticles.
Therefore, the series corresponding to the solution \eqref{7aa2},
\eqref{7aa2'} for $k_2\neq 0$ is not metastable. We shall assume in what follows that  $L_1\gg L_2$. Consider the Bogolyubov series corresponding to the flow along the capillary with velocity  $\hbar k_0/m$, where $k_0=2\pi(n_1/L_1,0,0)$. For the system of bosons under consideration the leading term of the asymptotics of the eigenvalues equals

\begin{equation}\label{Bogolubov Energy - k0}
N\l\frac{\hbar^2k_0^2}{2m}+\frac{V_0}{2}\r.
\end{equation}

Now let us assume that the relationship between
$L_1$ and $L_2$ is such that there exists a pair of vectors
$k_1$, $k=2\pi(0,n_2/L_2,n_3/L_2)$ for which the corresponding
value ~\eqref{11aa} is exactly equal to ~\eqref{Bogolubov Energy - k0}.
This means that there may be resonance between the superfluid states
of the Bogolyubov series and the nonflowing states of the metastable series
corresponding to ~\eqref{7aa2}, \eqref{7aa2'}. If $L_1$ is very large,
resonance is also possible for the case in which the value ~(\ref{11aa})
is close to ~(\ref{Bogolubov Energy - k0}) but does not necessarily
coincide with it.

The existence of such a resonance makes possible the passage from the
superfluid state to the metastable one, from which the system drops to the
lower energy level, which means that superfluidity is lost. The minimal energy
arising in the metastable series corresponds to the case  $k=2\pi(0,1/L_2,0)$ and,
according to formula ~(\ref{11aa}), equals

\begin{equation}
E_{\text{min}}=N\left(\frac{\hbar^2(2\pi)^2}{2mL_2^2}+\frac{V_0}2\right).
\label{15aa}
\end{equation}
Comparing ~(\ref{15aa}) with ~(\ref{Bogolubov Energy - k0}), we see that resonance
is impossible if the absolute value of the flow velocity $v$ is less than a certain number

\begin{equation}
|v|<v_c(L_2)\equiv\frac{2\pi\hbar}{mL_2} = \frac{h}{mL_2}.
\label{16aa}
\end{equation}

The quantity  $v_c(L_2)$ in the right-hand side of inequality ~(\ref{16aa})
increases as the cross section of the capillary  $L_2$ decreases. If $L_2$
is less than $2\pi\hbar/(m v_{cL})$, where  $v_{cL}$ is the critical Landau velocity
determined from ~(\ref{Bogolubov Spectrum}), then $v_c(L_2)$ is greater than
the critical Landau velocity and, accordingly, in this case superfluidity disappears when the Landau velocity is reached. However, when
 $L_2$ increases, the resonance between the flowing and metastable states arises
 before the Landau velocity is reached, and this explains the dependance of the critical velocity on the size of the cross section of the capillary.

Since the Landau curve and the Bogolyubov quasiparticles, as we showed in
 \cite{QuasiPart4}, do not change in the classical limit, it follows that, under the condition
 $v<h/(mL_2)$, where $L_2$  is the diameter of the nanotube, we have the following
 mathematical fact: the classical liquid in the nanotube must be superfluid.

Thus we have obtained a series of quasiparticles that differ from the Bogolyubov
series. This series corresponds to a ``standing wave'' across the tube, while the
Bogolyubov wave is a ``running wave'' along the tube.

The question of the dependence of these series on the temperature arises. As was shown by the author previously, the Bogolyubov series depends on the temperature,
which leads to the dependence of the Landau criterion (critical level) and to a new
phase transfer, which I called phase transfer of the zeroth kind.

It turns out that for the ``transverse series'' presented here and in
\cite{TMF_2005, RJ_2005}, the leading term of the asymptotics as $N \to \infty$
does not depend on the temperature. This follows from the fact the operator
$\wh{A}$ in the corresponding  $(L,A)$-pair which determines the dependence
on temperature in its leading term ``vanishes'' as  $N \to \infty$ in the sense
that  $\wh{A}^2 \equiv 0$.

Thus, for a sufficiently narrow tube, the phase transfer is determined by the Bogolyubov series only, so that here we have a phase transfer of the zeroth kind.

\textbf{Remark}

In our problem, the asymptotic decomposition is considered with respect to two
parameters: $N \to \infty$ and $h\to 0$ (the quasiclassical limit). How are they
related? In the 1947 Bogolyubov paper, there is no external potential, and so the problem is solved exactly. This means that, just as for linear equations with constant coefficients, the quasiclassical solution coincides with the exact one. But what if we introduce a potential? In that case, as we mentioned in the introduction, the quasiclassical approximation yields quasiparticles that preserve superfluidity (see
\cite{QuasiPart2} formula (24) and \cite{QuasiPart3}).

However, first of all,  the following conditions for the external potential must hold: in the classical problem of a self-consistent field, there must exist an invariant Lagrangian manifold up to ``atomic precision'', i.e., the potential must be ``absolutely precise'' and no noise of order $h$ is acceptable.

If this is not the case, the Bogolyubov spectrum in the three-dimensional case is
destroyed (see  \cite{Masl_Mish} and \cite{Book_Ultravt}) and the superfluidity
effect disappears.

Secondly, the potential must change very slowly, so that its change on a finite interval be less than a value of order $1/N$. Such a ``nanoprecision''and the
presence of such drastic conditions is possible only if nanotechnologies are used.

I surmise that a verifying experiment should be performed by using a neutral gas such as argon.

Let us note in conclusion that high temperature superfluidity, which was theoretically established by the author in 2004  \cite{FAN, DAN_2003, DAN_2004, DAN_2004', TMF_2004'}, is essentially the superfluidity of liquids in nanotubes.

I would like to express my gratitude to D.S.Golikov for recalculating and verifying all
the formulas.

\end{document}